\documentclass[aps,prd,groupedaddress,nofootinbib,longbibliography]{revtex4-1}

\usepackage[T1]{fontenc} 
\usepackage[utf8]{inputenc}
\usepackage{amsfonts,amsmath,amssymb}
\usepackage{mathtools}
\usepackage{nicefrac}
\usepackage{graphicx}
\usepackage{color}
\usepackage{soul} 
\usepackage{hyperref}
\usepackage{accents}
\usepackage[normalem]{ulem}
\usepackage{dcolumn}
\usepackage{mathrsfs}
\usepackage{cases}
\usepackage{braket}
\usepackage[utf8]{inputenc}
\usepackage{bbm}
 \usepackage{dutchcal}

\def\exd{\mathrm{d}}
\def\exD{\mathrm{D}}

\def\tr{\mathrm{tr}}
\def\d{\delta}
\def\o{\omega}

\def\G{\Gamma}

\def\cL{{\cal L}}

\def\Lie{{\cal L}}

\def\cH{\mathcal{H}}

\def\cA{{\cal A}}

\def\pa{\partial}

\def\bea{\begin{eqnarray}}
\def\eea{\end{eqnarray}}

\newcommand{\ubar}[1]{\underaccent{\bar}{#1}}

\newcommand{\h}[1]{\hat{#1}}

\def\ch{\mathcal{h}}
\def\D{\nabla}
\def\cA{\mathcal{A}}
\def\cU{\mathcal{U}}
\newlength{\dhatheight}
\newcommand{\hh}[1]{%
    \settoheight{\dhatheight}{\ensuremath{\hat{#1}}}%
    \addtolength{\dhatheight}{-0.35ex}%
    \hat{\vphantom{\rule{1pt}{\dhatheight}}%
    \smash{\hat{#1}}}}

\begin{document}


\title{Bogoliubov Transformation and Schrodinger Representation on Curved Space}


               \author{Musfar Muhamed Kozhikkal}
               \email[]{musfarmuhamed@gmail.com}
               \affiliation{Departament de F\'{i}sica Qu\`{a}ntica i Atrof\'{i}sica \\ Institut de Ci\`{e}ncies del Cosmos \\ Universidad de Barcelona, Spain.}
\author{Arif Mohd}
\email[]{arif7de@gmail.com}
\affiliation{Department of Physics \\ Aligarh Muslim University, India. 
               }


\date{\today}
\begin{abstract} 
  It is usually accepted that  quantum dynamics described by Schrodinger
  equation that determines the evolution  of states from one Cauchy surface to another 
  is unitary. However, it has been known for  some time that this expectation is not
  borne out in the conventional setting in which one envisages the dynamics on a fixed
  Hilbert space. Indeed it is not even true for linear quantum field theory
  on Minkowski space if the chosen Cauchy surfaces are not preserved by the flow of a timelike Killing vector.
  This issue was elegantly addressed and resolved by Agullo and Ashtekar who showed that in a general setting quantum dynamics in the Schrodinger picture
   does not take place in a fixed Hilbert space. Instead, it takes place on  a non-trivial bundle over time, the Hilbert bundle, whose fibre at a given time is a Hilbert space at that time.  
  In this article, we postulate a Schrodinger equation that incorporates the effect of change in vacuum during time evolution
  by including the Bogoliubov transformation explicitly in the Schrodinger equation. More precisely,
  for a linear (real) Klein-Gordon field on a globally hyperbolic spacetime we write down a Schrodinger equation
  that propagates states between arbitrary chosen Cauchy surfaces, thus describing the quantum dynamics on a Hilbert bundle.
  We show that this dynamics is unitary if a specific tensor on the canonical phase space satisfies the Hilbert-Schmidt condition. Generalized
  unitarity condition of  Agullo-Ashtekar follows quite naturally from our construction.
\end{abstract}

\maketitle
\tableofcontents

\section{Introduction}
\label{sec:intro}

We do not know of any physical principle or algorithm that selects the vacuum for quantum fields on
globally hyperbolic curved spacetimes. The choice of the vacuum is usually dictated by the symmetry of the background geometry,
 positivity of energy, and by requiring the renormalizability of the stress energy tensor. 
Indeed, even in Minkowski space there exist inequivalent representations of the operator algebra (see, for e.g., ref.~\cite{Wald:1995yp}). A well-known example
is the Rindler vs. Minkowski vacuum in flat space (the former is reducible, however). 
The representation that is  deemed physical is the one compatible with the symmetries of the Minkowski space.
Thus, for quantum field theory on Minkowski space a unique vacuum, hence a unique representation of the operator algebra, is picked out
 once the Poincare invariance of the vacuum is imposed.

In the usual approach to quantum field theory on curved space (\cite{Ashtekar:1975zn, Ashtekar:1980yw}), one starts by quantizing a classical 
phase space, which is usually taken to be the covariant phase space.  Classical theory also supplies the fields that are promoted to field operators in the quantum theory.
 The choice of vacuum is encoded in the choice of complex structure $J$ on the phase space. Complex structure is the new mathematical structure needed to go from the classical theory to the quantum theory. This leads to a one particle Hilbert space labeled by the points of phase space and the total Hilbert space is constructed as a Fock space built on top of the one particle Hilbert space. 
 Each choice of $J$ thus corresponds to a representation of the covariant operator algebra on a Hilbert space  labelled by $J$, $H_J$. This results in the covariant Heisenberg picture, in which the field operators are time dependent
and vacuum is a linear functional on the (commutator) algebra of field operators. The dynamical information is contained in the (representation of) operator algebra. The time dependent vacuum correlation functions of field operators  are the observables of the theory. 

In quantum mechanics, i.e., in the quantum description of finite number of degrees of freedom, the unitary equivalence of Heisenberg and
Schrodinger representation is well known. The celebrated Stone-von Neumann theorem assures that for finite number of particles
the representation of canonical commutation relations is essentially unique. The theorem however does not apply to field theories due
to the presence of infinite number of  degrees of freedom. 
One can thus ask if in the infinite dimensional setting, given a quantum field theory in the Heisenberg representation, can  a Schrodinger representation be constructed so that one has the notion of a state that evolves with time. In other words, the question is if in the field theory setting can the dynamics 
 be unitarily transferred from the operators to the states in the Hilbert space.  When the time evolution is along the timelike Killing vector field of the background spacetime then this can be done. I.e, when the spacetime is foliated in Cauchy slices orthogonal to the timelike Killing symmetry of the spacetime then the Heisenberg and Schrodinger representations are equivalent. That's why one doesn't ask this question in most of the quantum field theory textbooks.  However, many studies (see for e.g., refs.~\cite{Torre:1998eq,Helfer:1999uq,Helfer:1999ur,Helfer:1996my,Cortez:2013xt,Cortez:2012cf,Gomar:2012xn,Cortez:2012rj,GomezVergel:2007fd,BarberoG:2007frf,Corichi:2007ht,Cortez:2007hr,Corichi:2006zv,Torre:2002xt}.) reported  that this is not true in general. More precisely, in different settings in curved spacetimes, it was found in the mentioned studies that there is no unitary Schrodinger dynamics  between arbitrary Cauchy hypersurfaces (not necessarily orthogonal to the time translation Killing field of the background spacetime, or in the case that the background spacetime has no timelike Killing vector, e.g., in a cosmological setting) in a fixed Hilbert space. In particular, ref.~\cite{Torre:1998eq} showed that even in Minkowski space,  attempt to construct a unitary Schrodinger dynamics between arbitrary Cauchy slices fails. But there is no problem in the covariant quantization itself. It's only that one can not transfer the dynamics from the operator algebra to the states in a unitary fashion. 
 
 This perplexing state of affairs was addressed by Ashtekar and Agullo in an elegant paper, ~\cite{Agullo:2015qqa}.
These authors realized that the problem arises because of the insistence on describing the Schrodinger dynamics on a fixed Hilbert space. The approach of covariant quantization can be adapted to  canonical quantization as well. In this case, one again needs to choose a complex structure $J$, this time on the canonical phase space of the theory. The old unitarity condition, the one that is presented in textbooks, and the one that was shown to lead to the unitarity puzzle arises 
 by asking if there exists a unitary operator $U$ that evolves the state from initial time slice $t_i$ to a final time slice $t_f$ such that the following is true for any states $\ket{\Psi}$ and $\ket{\Psi'}$ $\in H_{J}$
 \begin{align}
\label{eq:oldU}
\bra{\Psi'} U(t_f, t_i)^{-1} \h{\mathcal{O}}(t_i) \,U(t_f, t_i) \ket{\Psi} = \bra{\Psi'}  \h{\mathcal{O}}(t_f)       \ket{\Psi},
\end{align}
where the field operators on the left and right hand side $\h{\mathcal{O}}$ are quantized with respect to $J$, i.e., they act on the fixed Hilbert space $H_J$. On the right hand side 
the field operator is evolved to the final time while the states are on the initial time. This is the Heisenberg representation of  quantum mechanics. On the left hand side the
 states are evolved to the final time slice while the operator is on the initial time. This is the Schrodinger representation of quantum mechanics. It turns out that such a unitary operator $U$ exists if and only if  $\left[J - EJE^{-1}\right]$ is Hilbert-Schmidt\footnote{We will give a proof of the Hilbert-Schmidt condition in the main text in sec.~\ref{subsec:hs}.}, where $E$ is the map implementing time evolution from $t_i$ to $t_f$ in the canonical phase space.  In several examples on curved space studied in refs.~\cite{Torre:1998eq,Helfer:1999uq,Helfer:1999ur,Helfer:1996my,Cortez:2013xt,Cortez:2012cf,Gomar:2012xn,Cortez:2012rj,GomezVergel:2007fd,BarberoG:2007frf,Corichi:2007ht,Cortez:2007hr,Corichi:2006zv,Torre:2002xt} this condition was shown to be violated. Even in simple cosmological examples that were first studied in refs.~\cite{Parker:1968mv, Parker:1969au, Parker:1971pt}, it was shown in ref.~\cite{Agullo:2015qqa} that the Hilbert-Schmidt condition is violated and hence such a unitary operator does not  exist. 

 The key insight of  ref.~\cite{Agullo:2015qqa} was that when the state evolves from one Cauchy slice to the other, one needs to take care of the evolution of the complex structure as well.
 Thus in the Schrodinger picture representation, the quantum dynamics does not take place on a fixed Hilbert space.  Ref.~\cite{Agullo:2015qqa} postulated a Generalized Unitarity condition, eq.~ \ref{eq:AAGU},  which appropriately takes care of the evolution of the complex structure. In this case, there is no {\it kinematical} unitary identification between the Hilbert spaces on the initial and final Cauchy slices,  $H_{J(t_i)}$ and 
$H_{J(t_f)}$, respectively. Instead, there is a {\it dynamical} unitary identification. Let's see this in a bit more detail. 

 Let $\ket{\Psi}$ and $\ket{\Psi'}$ be states  $\in H_{J(t_i)}$. Let $U(t_f, t_i): H_{J(t_i)} \rightarrow H_{J(t_f)}$, be the operator that maps $H_{J(t_i)}$ to $H_{J(t_f)}$.
 Let $\h{\mathcal{O}}$ be the operator quantized in the representation $J(t_i)$, thus it acts on states $\in H_{J(t_i)}$, and let $\hh{\mathcal{O}}$ be the same operator quantized in the representation $J(t_f)$, hence it acts on  $H_{J(t_f)}$. 
 The Generalized Unitarity condition of ref.~\cite{Agullo:2015qqa} is,
\begin{align}
\label{eq:AAGU}
\bra{\Psi'} U(t_f, t_i)^{-1} \hh{\mathcal{O}}(t_i) \,U(t_f, t_i) \ket{\Psi} = \bra{\Psi'}  \h{\mathcal{O}}(t_f)       \ket{\Psi}, 
\end{align}
where it should be noted that the hatted operator on the right hand side is acting at final time $t_f$ but is quantized in the representation $J(t_i)$ (thus acting on the Hilbert space
at $t_i$,  $H_{J(t_i)}$), while the double hatted operator on the left hand side is acting on the initial time $t_i$ but is quantized in the representation $J(t_f)$ (thus acting on the Hilbert space at $t_f$, $H_{J(t_f)}$). Therefore, the right hand side is the Heisenberg representation in which operator is evolving with time while the states and the representation space is fixed at the initial time. The left hand side is the Schrodinger representation in which the states and thus the representation space is evolving with time while the operator is acting at the initial time. The departure of ref.~\cite{Agullo:2015qqa} from the previous studies, and the reason why eq.~\ref{eq:AAGU} is called the {\it generalized} unitarity condition, is the fact that the Hilbert spaces $H_{J(t_i)}$ and $H_{J(t_f)}$ are in general not unitarily equivalent, in the sense that $\left[J(t_f) - J(t_i)\right]$ does not in general satisfy the Hilbert-Schmidt
 condition. However, eq.~\ref{eq:AAGU} leads to a different unitarity condition, namely, that  $\left[J(t_f) - EJ(t_i)E^{-1}\right]$ should be Hilbert-Schmidt, where $E$ is the map implementing time evolution from $t_i$ to $t_f$ in the canonical phase space. In particular, if $J(t_f)$ is the time evolution of $J(t_i)$ then we have that $J(t_f) = EJ(t_i)E^{-1}$ and the Hilbert-Schmidt condition is trivially satisfied and the operator $U$ is unitary. In the language of ref.~\cite{Agullo:2015qqa}, $U$ implements unitarily the dynamical automorphism of the operator algebra. 
 
 The goal of this paper is to explicitly construct the operator $U$  of eq.~\ref{eq:AAGU}. Our work is complementary to that of ref.~\cite{Agullo:2015qqa}. While in ref.~\cite{Agullo:2015qqa} the generalized unitarity condition was inspired by the 
 classical picture and the existence of the covariant quantum field theory  and then deriving the Schrodinger picture representation, our goal is to derive the generalized unitarity condition of ref.~\cite{Agullo:2015qqa}  by postulating the explicit quantum dynamics in the Schrodinger picture representation. We will use the ideas from geometric quantization in order to quantize the canonical phase space. We will also find it convenient to work in a fixed basis of the canonical phase space in a discrete notation, thus treating it as if it were finite dimensional, and  keeping our eye on the infinite dimensional limit in which the issue of unitarity arises. 
 
 The organization of this article is as follows. 
 In sec.~\ref{sec:classical} we review the construction of the canonical phase space for the real Klein-Gordon field theory on a globally hyperbolic spacetime. One goal in this
 section is to show that the symplectic form on the canonical phase space is translation invariant, a fact that we will use in the geometric quantization of the theory. Towards the
  end of this section we will also lay down our notation that we will follow in the later sections. In sec.~\ref{sec:quantum} we discuss the quantization procedure following refs.~\cite{Witten:1993ed, Axelrod:1989xt}.
   In sec.~\ref{subsec:gq}
  we review the geometric quantization, construction of  the Hilbert space and quantization of linear functions. In sec.~\ref{subsec:bog} we review the Bogoliubov transformation and the quantization of quadratic functions. In sec.~\ref{subsec:hs} we show how the Hilbert-Schmidt condition arises as the necessary condition for the unitary equivalence of two Hilbert spaces. 
In sec.~\ref{sec:sch} we present our Hamiltonian and the Schrodinger picture representation of quantum dynamics of the Klein-Gordon field. 
In sec.~\ref{sec:unitary} we make contact with ref.~\cite{Agullo:2015qqa} and derive their generalized unitarity condition from our Schrodinger equation. 
We conclude with a brief summary and outlook in sec.~\ref{sec:summary}. In app.~\ref{app:norm} we present the calculation for the normalization of wavefunction that is used in sec.~\ref{subsec:hs}.

\section{Scalar Field Theory: Classical}
\label{sec:classical}
In this section we sketch the construction of the canonical phase space of the classical free Klein-Gordon field theory. The purpose is to show that the canonical Hamiltonian
is a quadratic in fields and the symplectic structure is translation invariant on the phase space. This is important for our construction because we will be using this fact in the later sections.

Consider the free Klein-Gordon theory on a globally hyperbolic spacetime $M$ with a pseudo-Riemannian metric $g_{ab}$ described by the action
\begin{align}
S = -\frac{1}{2} \int \exd^4x \sqrt{-g} \left( g^{ab} \D_a \Phi \D_b \Phi + m^2 \Phi^2 \right),
\end{align}
where $g$ is the determinant of the metric, $g = det(g_{ab})$. 

Let $\Sigma$ be a Cauchy surface. Given a one parameter family of spacelike embeddings $\mathcal{I}_t$ of $\Sigma$ into $M$ we obtain a foliation of $M$ by a one parameter
family of hypersurfaces $\Sigma_t := \mathcal{I}_t  ( \Sigma )$ specified by a time function $t = constant$. Time evolution along the vector field  $t^a = \left(\frac{\partial}{\partial t}\right)^a$ can be decomposed in terms 
of unit timelike normal to $\Sigma_t$,  $n^a$ (where $n_a = \nabla_a t$), the lapse  function $N$, and the shift  vector $N^a$. In what follows we will put the shift to zero for simplicity. We then have the following relations:
\begin{align}
g^{ab} &= h^{ab} - n^a n^b, \\
t^a &= N n^a,
\end{align}
where $h_{ab}$ is the metric induced on $\Sigma_t$. The Klein-Gordon action can then be decomposed in a 3+1 form,
\begin{align}
S =  -\frac{1}{2} \int \exd t \int_{\Sigma_t} \exd^3 x N \sqrt{h} \left( h^{ab} D_a \Phi D_b \Phi - \frac{1}{N^2} (\Lie_t \Phi)^2 + m^2 \Phi^2 \right),
\end{align}
where $\Lie_t \Phi$ is the Lie derivative of $\Phi$ along the time evaluation vector field $t^a$, $\Lie_t \Phi = t^a \partial_a \Phi$, 
and $D_a$ is the covariant derivative operator induced on the hypersurface $\Sigma_t$ and is compatible with the induced metric $h_{ab}$.

Next, the canonical variables are defined as
\begin{subequations}
\begin{align}
\phi(x) &= \Phi(t,x) \vert_{\Sigma_t} \\
\pi(x) &= \frac{\delta S}{\delta \Lie_t \Phi} = \frac{1}{N} \Lie_t \Phi(t,x) \vert_{\Sigma_t}.
\end{align}
\end{subequations}
The canonical Hamiltonian density is obtained by a Lagrange transform and we get 
\begin{align}
\label{eq:Hclass}
\mathcal{H} = \frac{1}{2} N \left( h^{ab} D_a \phi D_b \phi + \pi^2 + m^2 \phi^2 \right),
\end{align}
This Hamiltonian generates the infinitesimal time evolution of the canonical variables $\phi$ and $\pi$ via the Poisson brackets,
\begin{subequations}
\label{eq:evol}
\begin{align}
\Lie_t \phi &= \{ \phi, \mathcal{H}\} = N \pi, \\
\Lie_t \pi  &= \{ \pi, \mathcal{H}\} = h^{ab} D_a N D_b \phi + N D^2 \phi - m^2 \phi.
\end{align}
\end{subequations}
We thus arrive at the canonical phase space of the Klein Gordon theory. The phase space $\Gamma$ is coordinatized by canonical variables $\phi(x)$ and $\pi(x)$,
i.e., a point on the phase space $\varphi$ is specified by its coordinates $\varphi = (\phi, \pi)$. The poisson bracket is seen as the  structure on the phase space which is given
 by the symplectic structure, which is a non-degenerate closed and exact two-form on the phase space,
 \begin{align}
 \omega(\delta_1, \delta_2) := \int_{\Sigma_t} \exd^3 x \sqrt{h} \left( \frac{\delta_1 }{\delta_1 \phi} \frac{\delta_2}{\delta_2 \pi} -  \frac{\delta_1 }{\delta_1 \pi} \frac{\delta_2}{\delta_2 \phi}\right).
 \end{align}
 
 Owing to the linearity of $\Gamma$, the symplectic structure can be pulled back from the tangent space of $\Gamma$ to $\Gamma$ thus giving $\Gamma$ the structure of a symplectic vector space with the symplectic form, which we again denote by $\omega$. Symplectic product between two phase space points $\varphi_1 = \left( \phi_1, \pi_1 \right)$ and $\varphi_2 = \left( \phi_2, \pi_2 \right)$ is given by
\begin{align}
\label{eq:symprod}
\omega(\varphi_1,\varphi_2) = \int_{\Sigma_t} \exd^3 x \sqrt{h} \left( \phi_1(x) \pi_2(x) - \phi_2(x) \pi_1(x)\right).
\end{align}
One reason to discuss this well-known classical setting is to emphasize that the symplectic structure is translation invariant in the sense that its components
do not depend on the phase space point. 

 Finite time evolution, from $t_i$ to $t_f$ on the canonical phase space can be obtained by integrating the Hamilton equations, eq.~\ref{eq:evol}. Equivalently, it can be written as a map\footnote{This is inverse of the map in ref.~\cite{Agullo:2015qqa}.}
 \begin{align}
 E_{t_f t_i} = \mathcal{I}_{t_f}^{-1} \mathcal{I}_{t_i}.
 \end{align}  
 The meaning of this equation is the following: for given functions $\phi(x)$ and $\pi(x)$ on $\Sigma$, first push them to $\Sigma_{t_i}$ using the map $\mathcal{I_{t_i}}$. Treating this as initial data on $\Sigma_{t_i}$, use the Klein Gordon equation $\left(\Box - m^2\right) \Phi(t,x) = 0$ to find the solution  $\Phi(t,x)$ corresponding to the initial data $(\phi, \pi)$. 
 Evaluate the solution  $\Phi(t,x)$  and its time derivative on the time slice  $\Sigma_{t_f}$ and push it back to $\Sigma$ using the map $\mathcal{I}_{t_f}^{-1}$. This process gives the finite time evaluation of the phase space point $\varphi$ at an initial time $t_i$ to a final time $t_f$ as  $ E_{t_f t_i} \cdot \varphi$.

 In what follows, we will work in a chosen basis for the canonical phase space. We will find it convenient to use a discrete notation and denote this basis as $\varphi^i$, where $i$ would in general stand for continuous indices like $x$. The discussion of this section can thus be summarized as follows. In the canonical description of the Klein-Gordon field we have a phase space $\Gamma$
whose coordinates are $\varphi^i$. Variation of any quantitiy with phase space point will be denoted by
 $\partial_i$ which means $\frac{\partial}{\partial \varphi^i}$. 
 $\Gamma$ is equipped with a non-degenerate closed and exact antisymmetric matrix, the canonical symplectic structure, with components $\omega_{ij}$. The inverse matrix of $\omega_{ij}$ will be denoted by $\omega^{ij}$, such that $\omega_{ij} \omega^{jk} = \delta^k_i$.  Note that components of the symplectic structure $\omega_{ij}$ are translation invariant in the sense that they do not depend upon the phase space points $\varphi^i$'s.  
 
 Infinitesimal time evolution on the phase space
is generated by the quadratic Hamiltonian which we write in component form as $\mathcal{H} = \frac12 H_{ij} \varphi^i \varphi^j$, where $H_{ij}$ is symmetric in its indices.
 The vector field on $\Gamma$ corresponding to any function $f$ on $\Gamma$  is given by $V_{f} = \omega^{-1}\exd f$.
  For our quadratic Hamiltonian this vector field has components  $V^i_\mathcal{H}= \omega^{ij} \partial_j \mathcal{H} = \omega^{ij} H_{jk} \varphi^k \equiv \left(\omega^{-1} H\right)^i_{~j} \varphi^j \equiv T^i_{~j} \varphi^j$. Thus infinitesimal time evolution of phase space is described by the matrix $T = \o^{-1} H$. Finite time evolution on $\Gamma$ is still given by the map  $E_{t_f t_i}$
 , i.e., $\varphi^i(t_f) = E^i_{~j}  \varphi^j(t_i)$.
 
\section{Scalar field theory: Quantum}
\label{sec:quantum}
In order to quantize the canonical phase space in the previous section we  follow the ideas from geometric quantization (\cite{Witten:1993ed, Axelrod:1989xt}) which is in fact closely related to the general construction of quantum field theory on curved spacetime in ref.~\cite{Ashtekar:1975zn} (see also ref.~\cite{Ashtekar:1980yw}) and used in ref.~\cite{Agullo:2015qqa}.  In both cases, one starts with the classical phase space and introduces a complex structure on it that is compatible with the symplectic structure in the sense discussed below. 

Accessible reviews of geometric quantization can be found in 
refs.~\cite{Nair:2016ufy,Woodhouse:1992de}.
Here we will follow refs.~\cite{Witten:1993ed, Axelrod:1989xt} to construct the quantum theory.  In sec.~\ref{subsec:gq} we will discuss the construction of the Hilbert space and the quantization of linear opeators. In sec.~\ref{subsec:bog} we describe the Bogoliubov transformation and the quantization of quadratic operators. In sec.~\ref{subsec:hs} we provide a proof of the Hilbert-Schmidt condition.

\subsection{Geometric Quantization}
\label{subsec:gq}

In geometric quantization one begins by first constructing a unitary line bundle on the phase space whose curvature is $-i \omega_{ij}$. If the symplectic structure is translation invariant 
then the covariant derivative on the sections of this bundle can be taken to be 
\begin{align}
\label{eq:exD}
\exD_i = \partial_i + \frac{i}{2} \omega_{ij} \varphi^j.
\end{align} 
Square integrable sections of this line bundle constitute the pre-quantum Hilbert space. Next one constructs a holomorphic structure on the pre-quantum line bundle by introducing the 
complex structure $J$ on the phase space such that: $J^2 = -1$ and $J$ is translation invariant in the sense that its components  $J^i_{~j}$ do not depend upon the point
 on the phase space where it is evaluated; $\omega_{ij}$ is compatible with $J$ in the sense that $\omega_{ij} J^i_{~p} J^j_{~q} = \omega_{pq}$; $J$ is positive in the sense that 
 it defines a positive metric $\text{g}_{ij}=\omega_{ik} J^k_{~j}$ on phase space. 
 
 The Hilbert space, i.e., the space of quantum states (wavefunctions), consists of  the square integrable sections of the pre-quantum Hilbert space which are holomorphic with respect
 to $J$, i.e., they satisfy the following equation
 \begin{align}
 (1 + i J)^k_{~i} \exD_k \Psi = 0.
 \end{align}
 Since this Hilbert space depends on the choice of complex structure $J$ it will be denoted by $H_J$. 
 
Following ref.~\cite{Witten:1993ed} it would be very convenient to introduce some notation that will streamline the calculations. Let us first introduce the projectors  that project on the 
holomorphic and anti-holomorphic sectors,
\begin{subequations}
\label{eq:proj}
\begin{align}
\frac12 (1 - i J)^i_{~j}  \,\,\,\,\,\,\,\,\,&\text{holomorphic}  , \\
  \frac12 (1 + i J)^i_{~j}\,\,\,\,\,\,\,\,\, &\text{anti-holomorphic}.
\end{align}
\end{subequations}
Projected components of vectors and one-forms will be represented by over/under-lined indices as :
\begin{subequations}
\begin{align}
V^{\ubar{i}} = \frac12 (1 - i J)^i_{~j} V^j \,\,\,\,\,\,\,\,\,&\text{holomorphic}, \\
V^{\bar{i}} = \frac12 (1 + i J)^i_{~j} V^j \,\,\,\,\,\,\,\,\, &\text{anti-holomorphic}, \\
V_{\ubar{i}} = \frac12 (1 - i J)^k_{~i} V_k \,\,\,\,\,\,\,\,\,&\text{holomorphic}, \\
V_{\bar{i}} = \frac12 (1 + i J)^k_{~i} V_k \,\,\,\,\,\,\,\,\, &\text{anti-holomorphic}.
\end{align}
\end{subequations}
The non-zero components of the complex structure are given by $J^{\ubar{i}}_{~\ubar{j}} = i\delta^{\ubar{i}}_{~\ubar{j}}$ and $J^{\bar{i}}_{~\bar{j}}= - i\delta^{\bar{i}}_{~\bar{j}}$. 
Furthermore, due to the condition of  compatibility between $\omega$ and $J$, the non-zero components of the symplectic structure are $\omega_{\ubar{i} \bar{j}}$ and
 $\omega_{\bar{i} \ubar{j}}$.

In this notation,  wavefunctions which lie in $H_J$ are holomorphic in the sense that their anti-holomorphic components vanish,
\begin{align}
\label{eq:holomorph}
\exD_{\bar{i}} \Psi = 0.
\end{align}
By using the definition of $\exD$ in eq.~\ref{eq:exD},  eq.~\ref{eq:holomorph} can be partially solved to get that any holomorphic wavefunction  $\Psi$ is of the form,
\begin{align}
\label{eq:wf}
\Psi = e^{-\frac{i}{2} \o_{\bar{i} \ubar{j}} \varphi^{\bar{i}} \varphi^{\ubar{j}}} g(\varphi),
\end{align}
where $g$ is a holomorphic function of coordinates in the ordinary sense, i.e., $\partial_{\bar{i}} g = 0, \forall i$. The wavefunction with $g=1$ is the vacuum wavefunction
in Hilbert space $H_J$,
\begin{align}
\label{eq:vac}
\Psi_{vac} = e^{-\frac{i}{2} \o_{\bar{i} \ubar{j}} \varphi^{\bar{i}} \varphi^{\ubar{j}}},
\end{align}
and is characterized by the fact that $\h{\varphi}^{\bar{i}} \Psi_{vac} = 0, \forall i$, where $\h{\varphi}$ is the operator to be defined in eq.~\ref{eq:fieldops}.

The inner-product on $H_J$ is given by 
\begin{align}
\label{eq:inprod}
\braket{\Psi_1, \Psi_2} = \int \left(\prod_{i=1}^{2N} \, d\varphi^{i} \right) \sqrt{\det{\left(\frac{\o}{2\pi}\right)}}\, \Psi_1^{\ast} \Psi_2,
\end{align}
where $2N$ is the dimension of phase space. In the infinite dimensional case, which is the case of interest for us,  we will take the limit $N \to \infty$ at the end of our calculations.

Next we introduce some operators on $H_J$. In geometric quantization there is a heuristic recipe to get operators from classical functions on phase space.
 Given a function $f$ on phase space, the pre-quantum operator $\hat{f}$ is constructed as 
\begin{align}
\hat{f} = - i V^i_f \exD_i + f,
\end{align}
where $V^i_f$ is the vector field generated by $f$ on the phase space, i.e., $V^i_f = \omega^{ij} \partial_j f$. This recipe is often to be augmented by the so called metaplectic corrections. 
For our purposes we will not need that technology. We will be interested only in quantizing the phase space functions which are at worst quadratic in phase space coordinates. Thus the technique given in refs.~\cite{Witten:1993ed, Axelrod:1989xt}  would be enough for our purpose. 
To begin with, we define the operators corresponding to linear functions on the phase space. The phase space coordinates $\varphi^i$ form a basis for these functions. Following the geometric quantization recipe, we get the corresponding operators as
\begin{align}
\hat{\varphi}^i = i \omega^{ij} \exD_j + \varphi^i,
\end{align}
where we evaluated the vector field corresponding to $\varphi^i$ as $V^j_{\varphi^i}= \omega^{jk} \partial_k \varphi^i = \omega^{ji}$. In particular, for holomorphic and
anti-holomorphic  components, using the fact that the wavefunctions are holomorphic (see eq.~\ref{eq:holomorph}) we get,
\begin{subequations}
\label{eq:fieldops}
\begin{align}
\hat{\varphi}^{\ubar{i}} &=  \varphi^{\ubar{i}}, \\
\hat{\varphi}^{\bar{i}} &= i \omega^{\bar{i} \ubar{j}} \exD_{\ubar{j}} + \varphi^{\bar{i}}.
\end{align}
\end{subequations}
Before we discuss the quantization prescription for the quadratic operators we need to understand how to take care of change in the complex structure. The reason is that we will
ultimately be interested in writing the Schrodinger equation, for which we need the Hamiltonian. However, in a general dynamical problem the complex structure will change with time.
Since the Hamiltonian is the infinitesimal generator of time evaluation, the finite time evolution should take us from the initial Hilbert space $H_{J(t_i)}$ to the 
final Hilbert space $H_{J(t_f)}$. We therefore turn to Bogoliubov transformation that implements the change of $J$ in quantum operators. 
This was constructed in Refs.~\cite{Axelrod:1989xt,Witten:1993ed} and we follow one of the constructions given in these references.

\subsection{Bogoliubov transformation and Quadratic operators}
\label{subsec:bog}
Let us consider a quadratic Hamiltonian $\ch = \frac12 h_{ij} \varphi^i \varphi^j$, where $h_{ij} = h_{ji}$. Vector field generated by $\ch$ on the phase space is
\begin{align}
V_{\ch}^i &= \o^{ij} \pa_j \ch  \nonumber  \\
&= \o^{ij} h_{jk} \varphi^k  \nonumber  \\
&\equiv \left(  \o^{-1}h\right)^i_{~k}\varphi^k.
\end{align}
Canonical transformation generated by $\ch$ is the flow of $V_{\ch}$ on the phase space obtained by the Lie-drag due to $V_{\ch}$ ,
\begin{align}
\d_{\ch} \phi^i &= \cL_{V_{\ch}} \varphi^i \nonumber \\
&= V_{\ch}^j \pa_j\varphi^i \nonumber \\
&\equiv \left( \o^{-1}h\right)^i_{~j} \varphi^j.
\end{align}
Change in the complex structure $J$ under this canonical transformation can be obtained by computing the Lie-derivative of $J$ along $V_{\ch}$. For translation invariant
complex structures (i.e., $\partial_k J^i_{~j} = 0$) we get
\begin{align}
 \d_{\ch} J^i_{~j} &= \cL_{V_{\ch}} J^i_{~j} \nonumber \\
&= \left[ J, \o^{-1}h\right]^i_{~j}
\end{align}
Now, choosing  $\o^{-1}h = -\frac12  \left( J \d J\right)$, and (since $J^2 = -1$) using that $\d J^2 = 0 = \left(J \d J + \d J J\right)$, we get 
\begin{align}
\d_{\ch} J^i_{~j} = \d J^i_{~j}.
\end{align}
Thus we see that any change in the complex structure $\d J$ can be obtained as a Hamiltonian flow on the phase space generated by the quadratic Hamiltonian 
$\ch = -\frac14 \left( \o J \d J \right)_{ij} \varphi^i \varphi^j$.  The vector field generated by $\ch$ is $V_{\ch} =  -\frac{1}{2} \left( J \d J \right)^i_{~j} \varphi^j$.

%

Follow the geometric quantization recipe for the pre-quantum operator corresponding to the classical function $\ch$ we get 
the pre-quantum operator $Pr(\ch)$,
\begin{align}
Pr(\ch) &= - i V_\ch^i \exD_i + \ch \nonumber \\
&= \frac{i}{2}  \left(J \d J \right)^i_{~j} \varphi^j \exD_i - \frac{1}{4} \left( \o J \d J \right)_{ij} \varphi^i \varphi^j.
\end{align} 
Note that the pre-quantum operator is a first-order differential operator. Following ref.~\cite{Axelrod:1989xt} (see ref.~\cite{Witten:1993ed} for another approach) let's write another operator corresponding to $\ch$, 
denoted by $\h{\ch}$, which is obtained by simply putting hats on $\varphi$'s,
\begin{align}
\h{\ch} &= \frac12 h_{ij} \h{\varphi}^i \h{\varphi}^j \\
&= \frac12 h_{ij} \left(i \o^{im} \exD_m + \varphi^i \right) \left(i \o^{jn} \exD_n + \varphi^j\right). 
\end{align}
where we don't need to symmetrize since $h_{ij}$ is already symmetric.  Note that $\h{\ch}$ does not depend upon the complex structure $J$. 
Expanding the right hand side we get that 
\begin{align}
\label{eq:bog}
\h{\ch} = Pr(\ch) - \frac{1}{4} \left( J \d J \o^{-1}\right)^{ij} \exD_i \exD_j.
\end{align}
The meaning of this equation is as follows. The operator $\h\ch$ on the left hand side does not depend on $J$. The first term on the right hand side, $Pr(\ch)$, implements 
the Bogoliubov transformation. Therefore, the second term on the right hand side is interpreted as canceling the change of complex structure.  Since $\exD$'s act on wavefunctions $\Psi$ which are holomorphic
with respect to $J$, and using that  $J^{\ubar{i}}_{~ \ubar{j}} = i \d^{\ubar{i}}_{~ \ubar{j}}$, the second term in eq.\ref{eq:bog} can be written as 
\begin{align}
\label{eq:connection}
\cA :=  \frac14 \left( \d J \o^{-1}\right)^{\ubar{i}\ubar{j}}  \exD_ {\ubar{i}}\exD_{\ubar{j}},
\end{align}
where the symbol $\cA$ comes from the fact that $\cA$ is the connection on a bundle. Indeed, in the approach of ref.~\cite{Axelrod:1989xt}, one constructs a bundle whose base space
is the space of $J$'s and whose fibre at $J$ is the Hilbert space $H_J$. Then $\cA$ denotes the connection that is to be used for transporting a state from the fibre $H_{J}$ at
$J$ to the fibre  $H_{J'}$ at $J'$. For finite dimensional systems this connection is unitary, in the sense that the states are mapped by this transport in a unitary fashion. In 
the infinite dimensional case though this connection will generically be non-unitary. However, we will see that this is not a problem. The map that would take us from the Hilbert space at
initial time $t_i$ to a final time $t_f$ will in fact be unitary, thus demonstrating the implementation of ``dynamical automorphism'' of operator algebra as a unitary transformation between the two Hilbert spaces as described in ref.~\cite{Agullo:2015qqa}. 

This leads us to the prescription of ref.~\cite{Axelrod:1989xt} for the quantization of classical 
functions quadratic in phase space coordinates. For such a function $f$, the quantum operator $\h{f}$ is given by
\begin{align}
\label{eq:f}
\h{f} &=  Pr(f) +  \frac{i}{4} \left( \d_f J \o^{-1}\right)^{\ubar{i}\ubar{j}}  \exD_ {\ubar{i}}\exD_{\ubar{j}},
\end{align}
where $Pr(f) = - i V_f^i \exD_i + f$, and $\d_f J$ is the change in complex structure along the flow of the vector field $V_f$ on the phase space, i.e., $\d J = \Lie_{V_f} J$. 
It can be checked that this quantization scheme respects the Poisson bracket structure,
\begin{align}
[\h{f}, \h{\varphi}^k ] = - i \widehat{\left\{ f, \phi^k \right\}}.
\end{align}

\subsection{Unitarity of $\cA$: The Hilbert-Schmidt condition}
\label{subsec:hs}

Consider the phase space $\G$ for a free scalar field theory. Let's construct two Hilbert spaces $H_1$ and ${H}_2$ corresponding to two complex structures 
$J_1$ and $J_2$, respectively, on $\G$. Then $H_1$ and $H_2$ are unitarily equivalent iff the following condition is satisfied
\begin{align}
\label{eq:hs}
\tr \,\, |J_1 - J_2|^2 < \infty,
\end{align}
where the trace is over the phase space\footnote{which is equivalently the trace over one-particle Hilbert corresponding to either $H_1$ or $H_2$ after Cauchy completion.}. The inequality in eq.~\ref{eq:hs}  is called the Hilbert-Schmidt condition. 
Here we provide  a physicist's derivation of the Hilbert-Schmidt condition  following ref.~\cite{Ashtekar:1980yw}. 

We could use the connection to propagate the vacuum
 wavefunction (see, eq.~\ref{eq:vac}) with respect to $J_1$ and checking if the resulting wavefunction $\in H_{J_2}$, i.e., if it  is normalizable in $H_{J_2}$. The propagation is
 implemented by the exponentiated version of $\cA$, which is defined as $\cU = \mathcal{P} e^{\int \cA}$ (path integrated exponential) that maps the state from $H_{J_1}$ to $H_{J_2}$. However, it is easier to follow another route as in ref.~\cite{Ashtekar:1980yw}. To this end,
 we first compute that commutator of the connection with a field operator $\h{\varphi}^i$ and we get,
 \begin{align}
 \left[ \cA, \h{\varphi}^i \right] = 0.
 \end{align}
This means that $U$ is such that
 \begin{align}
 U^{-1} \hh{\varphi}^i U =  \h{\varphi}^i,
 \end{align}
 where, since $U : H_{J_1} \rightarrow H_{J_2}$, we have that the double hatted operator  on the left hand side is quantized with respect to $J_2$ and the one on the right hand side is quantized with respect to $J_1$. In particular, 
 \begin{align}
  U^{-1} \left(1+ iJ_1\right) \hh{\varphi} U =  \left(1+ iJ_1\right) \h{\varphi},
 \end{align}
 where we have omitted the indices for brevity. Operating the right hand side on the vacuum in $H_{J_1}$, $\Psi_{vac}$ (see eq.~\ref{eq:vac}), we get zero. Let $\psi = U \Psi_{vac}$. Then we have that
 \begin{align}
  \left(1+ iJ_1\right)^i_{~j} \h{\varphi}^j \Psi_{vac} &= 0, \nonumber\\
  &\implies \left[ \left(1+ iJ_1\right)^i_{~\bar{j}} \hh{\varphi}^{\bar{j}} + \left(1+ iJ_1\right)^i_{~\ubar{j}} \hh{\varphi}^{\ubar{j}} \right] \psi = 0,\nonumber\\
  &\implies \left[ \left(J_2+ J_1\right)^i_{~\bar{j}} \hh{\varphi}^{\bar{j}} - \left(J_2 - J_1\right)^i_{~\ubar{j}} \hh{\varphi}^{\ubar{j}} \right] \psi = 0, \nonumber\\
&\implies \left[  \hh{\varphi}^{\bar{i}} - \chi^{\bar{i}}_{~\ubar{j}} \hh{\varphi}^{\ubar{j}}\right] \psi  = 0, \label{eq:chi}
 \end{align}
 where $\hh{\phi}$ again signifies that the indices of the operator are projected with respect to $J_2$, and $\chi$ is the operator $\chi:= (J_2 + J_1)^{-1}(J_2 - J_1)$ . Next we substitute the operator definitions (see eq.~\ref{eq:fieldops}) to get the differential equation
 \begin{align}
 \left[ i \omega^{\bar{i}\ubar{j}} \exD_{\ubar{j}} + \varphi^{\bar{i}} -  \chi^{\bar{i}}_{~\ubar{j}} {\varphi}^{\ubar{j}} \right] \psi = 0.
 \end{align}
 This equation can be simplified by first expressing $\psi$ as  $\psi = e^{-\frac{i}{2} \omega_{\bar{i}\ubar{j}}\varphi^{\bar{i}} \varphi^{\ubar{j}}} g(\varphi)$, for some holomorphic
 (with respect to $J_2$) function $g$,
 (see eq.~\ref{eq:wf}), which gives a differential equation for $g$ as 
 \begin{align}
 \left[ i \omega^{\bar{i}\ubar{j}} \pa_{\ubar{j}} -  \chi^{\bar{i}}_{~\ubar{j}} {\varphi}^{\ubar{j}}\right] g  = 0,
 \end{align}
 which is easily solved to get 
 \begin{align}
 g(\varphi) = e^ {\frac{i}{2} \o_{\bar{i}\ubar{j}}\chi^{\bar{i}}_{~\ubar{k}}\varphi^{\ubar{k}} \varphi^{\ubar{j}}}
 \end{align}
 Therefore the vacuum $\Psi_{vac} \in H_{J_1}$ propagated to $H_{J_2}$ yields the state  $\psi = U \Psi_{vac}$,
 \begin{align}
 \psi =  \mathscr{N} \, e^{-\frac{i}{2} \o_{\bar{i}\ubar{j}}\varphi^{\bar{i}} \varphi^{\ubar{j}}} e^ {\frac{i}{2} \o_{\bar{i}\ubar{j}}\chi^{\bar{i}}_{~\ubar{k}}\varphi^{\ubar{k}} \varphi^{\ubar{j}}},
 \end{align}
 where $\mathscr{N}$ is yet-to-be-determined normalization constant. $\psi$ would be a legitimate state and would lie in $H_{J_2}$ if $\mathscr{N}$ is  finite. 
 Calculation of the normalization factor is given in the appendix~\ref{app:norm}, where we show that  the condition for the finiteness of $\mathscr{N}$ is that 
$\tr\, \chi^2 := \chi^{\bar{i}}_{~\ubar{j}} \chi^{\ubar{j}}_{~\bar{i}}$ should be finite, which translates to the Hilbert-Schmidt condition
 \begin{align}
 \label{eq:hsc}
 \left(J_2 - J_1\right)^{\bar{i}}_{~\ubar{k}}  \left(J_2 - J_1\right)^{\ubar{k}}_{~\bar{i}} < \infty. 
 \end{align}
 
\section{Hamiltonian Evolution and The Schrodinger equation}
\label{sec:sch}
Let's now consider a linear system whose time evaluation is given by a Hamiltonian quadratic in coordinates, as it is the case for us (see eq.~\ref{eq:Hclass}), 
\begin{align}
\cH = \frac12 H_{ij} \varphi^i \varphi^j.
\end{align}
From the quantization prescription in eq.~\ref{eq:f} for such a  function we have the corresponding operator, 
\begin{align}
\label{eq:Hop}
 \h{\cH} =  Pr(H) + \frac{i}{4} \left( \d_HJ \o^{-1}\right)^{\ubar{i}\ubar{j}}  \exD_ {\ubar{i}}\exD_{\ubar{j}},
\end{align}
where $\d_H J$ is the change in complex structure due to flow generated by $H$ on phase space. While one can choose the complex structure at 
each time independently, we will consider the natural time evolution of $J$. This means that $\d_H J = \Lie_{V_H} J = \left[J, T \right]$, where, recall
from the last paragraph of sec.~\ref{sec:classical}, $T$ is the matrix describing the infinitesimal time evolution of the phase space points $(\dot{\varphi}^i = T^i_{~j} \varphi^j)$.
 The reason
that we keep only the natural time evolution of the initial $J$ in our Hamiltonian is that once we have the complete picture of time evolution of wavefunctions
in this case, then we can simply do a Bogoliubov transformation to any other complex structure that we decide to choose on the final time slice. More discussion about this
appears at the end of sec.~\ref{sec:unitary}.

 It can be checked that with this prescription the quantization respects the Poisson bracket,
\begin{align}
\left[\h{\cH}, \h{\varphi}^k \right] = - i \widehat{\left\{ \cH, \varphi^k \right\}}.
\end{align}
Importantly, our Hamiltonian respects the holomorphicity of wavefunctions. This is so because we can check that
\begin{align}
\label{eq:comHD}
\left[\h{\cH}, \exD_{\bar{i}} \right]  = 0.
\end{align}
To see that this condition is necessary for preserving the holomorphicity of wavefunctions let us see the implication of $\d\left(\exD_{\bar i} \Psi \right) = 0 $,
\begin{align}
\d \left( \exD_{\bar i} \Psi \right) &= \d \exD_{\bar i} \cdot \Psi + \exD_{\bar i} \, \left(\d \Psi\right) \nonumber \\
 &= i \left[\h{\cH}, \exD_{\bar{i}} \right] \Psi +  \exD_{\bar{i}} \left(-i \h{\cH} \Psi \right) \nonumber \\
 & =  i \left[\h{\cH}, \exD_{\bar{i}} \right] \Psi - i \left[\exD_{\bar{i}},  \h{\cH} \right] \Psi \nonumber \\
 & = 2i \left[\h{\cH}, \exD_{\bar{i}} \right] \Psi,
\end{align}
where in going to the third equality from the second we used the holomorphicity of $\Psi$. Therefore, $\d \left( \exD_{\bar i} \Psi \right) = 0 $ $\implies$
 $ \left[\h{\cH}, \exD_{\bar{i}} \right] = 0$. Thus,  eq.~\ref{eq:comHD} says that our Hamiltonian keeps the holomorphic wavefunctions holomorphic. 

Our proposal for the Schrodinger equation, i.e., the equation describing the time evolution of the wavefunction is then 
\begin{align}
\label{eq:sch}
i \frac{\pa\psi}{ \pa t} = \h{\cH} \psi,
\end{align}
where $\h{\cH}$ is as given in eq.~\ref{eq:Hop} and includes the connection term. Eq.~\ref{eq:sch} is the key equation of this paper.

From our Schrodinger equation in eq.~\ref{eq:sch} we have the following finite time evolution of a quantum state,
\begin{align}
\Psi(t_f) = U(t_f, t_i) \Psi(t_i),
\end{align}
where,
\begin{align}
\label{eq:ourU}
U(t_f, t_i) = \mathcal{T} \left( e^{-i \int_{t_i}^{t_f}\exd t \,\h{\cH}} \right),
\end{align}
where $\mathcal{T}$ is the symbol for time ordering dictated by our foliation of spacetime. The commutation relation in eq.~\ref{eq:comHD}  says that our Hamiltonian at time $t$, which depends upon the complex structure $J$ at time $t$, preserves the holomorphicity of  wavefunctions at time $t$. This implies that the exponentiation yielding the operator for finite time evolution $U(t_f, t_i)$ maps the wavefunctions at time $t_i$ that are holomorphic with respect to $J(t_i)$ to the wavefunctions at time $t_f$ that are holomorphic with respect to $J(t_f)$.
Thus $U(t_f, t_i)$ maps the Hilbert space at time $t_i$, $H_{J(t_i)}$ to the Hilbert space at time $t_f$, $H_{J(t_f)}$,
\begin{align}
U(t_f, t_i) :  H_{J{(t_i)}} \rightarrow H_{J{(t_f)}}.
\end{align}

The question now is, under what condition is $U$ a unitary operator? In order to answer this, we could just propagate the vacuum state $\Psi_{vac}$ in $H_{J_{t_i}}$ using $U$ and
check for the finiteness of the norm of the  evolved state $U(t_f, t_i) \Psi_{vac}$ to see if it lies in $H_{J(t_f)}$. But it is easier to follow along the lines of analysis in 
sec.~\ref{subsec:hs}. In the next section we will first derive the Generalized Unitarity condition of ref.~\cite{Agullo:2015qqa} and then use it to prove the unitarity of $U$.

\section{Unitarity of Time Evolution}
\label{sec:unitary}
Let us begin by collecting at one place our expression of the quantized Hamiltonian operator,
\begin{align}
\label{eq:quH}
\h{\cH} = Pr{(H)} + \frac{i}{4} \left(\d_H J \o^{-1}\right)^{\ubar{i} \ubar{j}} \exD_{\ubar{i}} \exD_{\ubar{j}}, 
\end{align}
where,
\begin{align}
Pr(H) = - i V_H^{\ubar{i}} \exD_{\ubar{i}} + H &= - i T^{\ubar{i}}_{~k} \varphi^k \exD_{\ubar{i}} + \frac12 (\o T)_{ij} \varphi^i \varphi^j, \\
\frac{i}{4} \left(\d_H J \o^{-1}\right)^{\ubar{i} \ubar{j}} \exD_{\ubar{i}} \exD_{\ubar{j}} &= \frac12 (T\o^{-1})^{\ubar{i} \ubar{j}} \exD_{\ubar{i}} \exD_{\ubar{j}},
\end{align}
where we have used that $\d_H J = \Lie_{V_H}J= [J, T]$. Furthermore, the following commutation relations follow,
\begin{subequations}
\begin{align}
i \left[\h{\cH}, \h{\varphi}^{\ubar{i}}\right]  &= T^{\ubar{i}}_{~k} \h{\varphi}^k, \\
i \left[\h{\cH}, \h{\varphi}^{\bar{i}} \right] &= T^{\bar{i}}_{~{k}} \h{\varphi}^{{k}}.
\end{align}
\end{subequations}

Consider an initial state  $\Psi \in H_{J(t_i)}$ and its time evolved image\footnote{There is a slight abuse of notation here, for we don't know yet if the evolved state is $\in H_{J(t_f)}$.} $U(t_f, t_i) \Psi \in  H_{J(t_f)}$. Now an operator that acts on the time evolved
state is ${\hh\varphi^{\ubar{k}}}$. Notice that here $\ubar{k}$ means that the index $k$ is holomorphically projected with respect to $J(t_f)$, i.e,  the operator is quantized with respect to $J(t_f)$ since it acts on the Hilbert space $H_{J(t_f)}$. The latter fact is emphasized by the double-hatted notation. Similarly,  ${\hh\varphi^{\bar{k}}}$ also acts on $H_{J(t_f)}$ .  Note that we want to interpret these as Schrodinger picture operators so we are inserting them still at the initial time $t_i$, but they act on the Hilbert space at $t_f$. 

For any  states $\ket{\Psi'}$ and $\ket{\Psi} \in  H_{J(t_i)} $,  infinitesimal form of  the operator insertion in $\bra{\Psi'} U(t_f, t_i)^{-1} {\hh\varphi^{\ubar{i}}} \,U(t_f, t_i) \ket{\Psi}$ and
 $\bra{\Psi'} U(t_f, t_i)^{-1} {\hh\varphi^{\bar{i}}} \,U(t_f, t_i) \ket{\Psi}$, in the limit that $t_f \rightarrow t_i$ can be calculated to be,
 \begin{subequations}
\begin{align}
i \left[\h{\cH}, \h{\varphi}^{\ubar{i}} \right] -  \frac{i}{2} \d_H J^i_{~j} \h{\varphi}^j &= T^i_{~\ubar{j}} \h{\varphi}^{\ubar{j}}, \\
i \left[\h{\cH}, \h{\varphi}^{\bar{i}} \right] +  \frac{i}{2} \d_H J^i_{~j} \h{\varphi}^j &= T^i_{~\bar{j}} \h{\varphi}^{\bar{j}}, 
\end{align}
\end{subequations}
respectively. From these infinitesimal forms we deduce the corresponding finite forms,
\begin{subequations}
\begin{align}
\bra{\Psi'} U(t_f, t_i)^{-1} {\hh\varphi^{\ubar{i}}} \,U(t_f, t_i) \ket{\Psi} = \bra{\Psi'}  E^i_{~j}\left(\frac{1-iJ(t_f)}{2}  \right)^j_{~k} \h{\varphi}^k        \ket{\Psi}, \\
\bra{\Psi'} U(t_f, t_i)^{-1} {\hh\varphi^{\bar{i}}} \,U(t_f, t_i) \ket{\Psi} = \bra{\Psi'}  E^i_{~j}\left(\frac{1+iJ(t_f)}{2}  \right)^j_{~k} \h{\varphi}^k        \ket{\Psi},
\end{align}
\end{subequations}
respectively. Here $E^i_{~j}$ is the classical time evaluation on the canonical phase space $\Gamma$,  $\left(\varphi^i(t_f) = E^i_{~j}  \varphi^j(t_i) \right)$ , and the operator on the right hand side is quantized with respect to $J(t_i)$ since it acts on the states in the initial Hilbert space $H_{J(t_i)}$. Adding these two equations we get,
\begin{align}
\label{eq:GU}
\bra{\Psi'} U(t_f, t_i)^{-1} {\hh\varphi^{{i}}} \,U(t_f, t_i) \ket{\Psi} = \bra{\Psi'}  E^i_{~j} \h{\varphi}^j        \ket{\Psi}, 
\end{align}
where again the operator on the right hand side is quantized with respect to $J(t_i)$ while the one on the left hand side is quantized with respect to $J(t_f)$.
Eq.~\ref{eq:GU} is precisely the  Generalized Unitarity condition (eq.~\ref{eq:AAGU}) proposed in ref.~\cite{Agullo:2015qqa}.
If the operator $U$ is unitary  then eq.~\ref{eq:GU} establishes the equivalence of the Heisenberg and Schrodinger picture in quantum field theory for arbitrary foliation of the background spacetime. 

We turn to the  analysis of unitarity of $U$ now. The strategy would be the same as in sec.~\ref{subsec:hs}. We project the operator on the right hand side to its anti-holomorphic components and act on the vacuum state in the initial Hilbert space. We have the following operator relation from eq.~\ref{eq:GU}
 \begin{align}
 U(t_f, t_i)^{-1} {\hh\varphi^{{i}}} \,U(t_f, t_i) =  E^i_{~j} \h{\varphi}^j.
 \end{align}
 To get the anti-holomorphic components on the right hand side we perform the following manipulations (suppressing the indices for  a little bit),
 \begin{align}
 E\left(1+iJ(t_i)\right)\h{\varphi} &= \left(1+ i E J(t_i) E^{-1}\right) E \h{\varphi} \nonumber\\
 &=  U(t_f, t_i)^{-1} \left(1+ i E J(t_i) E^{-1}\right) {\hh\varphi} \,U(t_f, t_i).
 \end{align}
 Now applying $E\left(1+iJ(t_i)\right)\h{\varphi} $ to $\Psi_{vac} \in H_{J(t_i)}$ we get zero. Let $U(t_f, t_i) \Psi_{vac} = \psi \in H_{J(t_f)}$. We then get
 \begin{align}
 &\left[\left(1+ i E J(t_i) E^{-1}\right) {\hh\varphi} \right] \psi = 0.  \nonumber \\
 \implies & \left[ \left(1+ i E J(t_i) E^{-1}\right)^i_{~\bar{j}} {\hh\varphi^{\bar{j}}} +  \left(1+ i E J(t_i) E^{-1})\right)^i_{~\ubar{j}} {\hh\varphi^{\ubar{j}}} \right] \psi = 0,  \nonumber \\
 \implies & \left[ \left(J(t_f) +  E J(t_i) E^{-1}\right)^i_{~\bar{j}} {\hh\varphi^{\bar{j}}} -  \left( J(t_f) -  E J(t_i) E^{-1})\right)^i_{~\ubar{j}} {\hh\varphi^{\ubar{j}}} \right] \psi = 0, \nonumber\\
 \implies & \left[  \hh{\varphi}^{\bar{i}} - \chi^{\bar{i}}_{~\ubar{j}} \hh{\varphi}^{\ubar{j}}\right] \psi  = 0, \label{eq:reph}
 \end{align}
where $\chi$ is the operator $\chi:=  \left(J(t_f) +  E J(t_i) E^{-1}\right)^{-1} \left(J(t_f) -  E J(t_i) E^{-1}\right)$. In the second equality above we have restored the indices and decomposed the operator in terms of its holomorphic and anti-holomorphic components with respect to $J(t_f)$. In the third equality we have used the fact that $J(t_f)^i_{~\bar{j}} {\hh\varphi^{\bar{j}}} = - i {\hh\varphi^{\bar{j}}}$ and similarly for the holomorphic component. But eq.~\ref{eq:reph} is the same as eq.~\ref{eq:chi} with the identification 
$J_2 = J(t_f)$ and $J_1 = E J(t_i) E^{-1}$. Therefore the result eq.~\ref{eq:hsc} of sec.~\ref{subsec:hs} can be imported as is. We thus have the conclusion that operator $U(t_f, t_i)$ is unitary iff
 \begin{align}
 \label{eq:ghsc}
   \left[ J(t_f) -  E J(t_i) E^{-1}\right]^{\bar{i}}_{~\ubar{k}}   \left[ J(t_f) -  E J(t_i) E^{-1}\right]^{\ubar{k}}_{~\bar{i}} < \infty.
 \end{align}
In our case, since our the complex structure $J(t_f)$ is not specified separately but is dictated by the time evolution of the initial complex structure  $J(t_f)$, we have that 
 $J(t_f) = E J(t_i) E^{-1}$ and thus the Hilbert-Schmidt condition is trivially satisfied. This matches with the conclusion of ref.~\cite{Agullo:2015qqa}.
  Therefore, our evolution operator $U$ is always unitary. Note that in the finite dimensional case there is nothing to check because there will be finite number of terms
  in eq.~\ref{eq:ghsc} and $U$ will be unitary. However, in the infinite dimensional case eq.~\ref{eq:ghsc} is a non-trivial condition for ensuring the unitarity of Schrodinger dynamics. 
  
   What if one insists on specifying another 
 complex structure on the final slice, say, $J_{new}$? In that case our strategy would be to  first use our Schrodinger equation~\ref{eq:sch} to propagate the state till the final time slice to get an element of  $H_{J(t_f)}$  and then do a Bogoliubov transformation to get a state in $H_{J_{new}}$. From eq.~\ref{eq:hsc} in sec.~\ref{subsec:hs} the latter will be unitary
  iff $\left[J_{new} - J(t_f) \right]$ is Hilbert-Schmidt, i.e., iff
  \begin{align}
     \left[ J_{new}-J(t_f)\right]^{\bar{i}}_{~\ubar{k}}   \left[ J_{new} -  J(t_f)\right]^{\ubar{k}}_{~\bar{i}} < \infty,
  \end{align}
  where (anti-)holomorphic projections are with respect to $J_{new}$.


\section{Summary and Outlook}
\label{sec:summary}
Quantum field theory on curved spacetime is usually studied in the covariant picture. In this picture the covariant operator algebra is given on the spacetime. The choice of vacuum is equivalent to the choice of complex structure on the covariant phase space. Once one chooses this complex structure one can construct the wavefunctions which are holomorphic functions with respect to this complex structure. Therefore the Hilbert space is labeled by this complex structure and provides the representation space for the representation of the covariant  operator algebra. This is done either in terms of creation/annihilation operators or by geometric quantization as discussed in this paper. In practical applications one specifies the complex structure by choosing  a basis of the solutions of the linear field equation and decomposing them in terms of positive and negative frequency. This induces a holomorphic structure on the covariant phase space. 
 In this approach dynamics is carried by the operators and state is considered fixed. Correlation functions of operators in this state are the observables of the theory. This Heisenberg picture of quantum dynamics is 
 sufficient and complete. However, if one attempts to transfer the time dependence from the operators to the states thus constructing a Schrodinger picture of quantum dynamics one runs into the problem noted in refs.~\cite{Torre:1998eq,Helfer:1999uq,Helfer:1999ur,Helfer:1996my,Cortez:2013xt,Cortez:2012cf,Gomar:2012xn,Cortez:2012rj,GomezVergel:2007fd,BarberoG:2007frf,Corichi:2007ht,Cortez:2007hr,Corichi:2006zv,Torre:2002xt} that the Schrodinger dynamics turns out to be non-unitary. This issue was resolved by Agullo and Ashtekar in 
 ref.~\cite{Agullo:2015qqa} who realized that the problem arises because one is trying to construct  the Schrodinger dynamics on a fixed Hilbert space. If one insists on a fixed Hilbert space then eq.~\ref{eq:oldU} indeed leads to the conclusion that the operator $U$ is not unitary. However, if one allows the Hilbert space to change during time evolution then
  ref.~\cite{Agullo:2015qqa} postulated a generalized unitarity condition stated in  eq.~\ref{eq:AAGU} which leads to a unitary $U$.  The goal of our paper is to understand the operator
  $U$ that appears in eq.~\ref{eq:AAGU}. In fact our objective is complementary to that of  eq.~\ref{eq:AAGU}. We postulated the Schrodinger dynamics directly by constructing the Hamiltonian operator in eq.~\ref{eq:quH} that evolves the states infinitesimally. To this end we used the Bogoliubov transformation implemented by the connection in eq.~\ref{eq:connection} first  constructed in refs.~\cite{Witten:1993ed, Axelrod:1989xt}. Operator $U$ is then the time-ordered exponential of the Hamiltonian, eq.~\ref{eq:ourU}.  
 We then derived the generalized unitarity condition of  ref.~\cite{Agullo:2015qqa}  and the Hilbert-Schmidt condition that follows from it in sec.~\ref{sec:unitary}. Given that we postulated the Schrodinger equation, one might wonder why did we have to derive the generalized unitarity condition before checking the unitarity of the Schrodinger dynamics. Indeed, it would be nice to directly evolve the vacuum wavefunction in eq.~\ref{eq:vac} using our $U$, but we found it simpler to first derive the generalized unitarity condition.  
 
 Note that the condition for unitarity of our $U$ is that  $\left[ J(t_f) -  E J(t_i) E^{-1}\right]$ be Hilbert-Schmidt. This is different from the old unitarity condition which insisted on the dynamics on a fixed Hilbert space $H_J$ and which leads to the condition that  $\left[ J -  E J E^{-1}\right]$ be Hilbert-Schmidt. Moreover, the condition is {\it not} that
  $\left[ J(t_f) -  J(t_i)\right]$  be Hilbert-Schmidt, which would be the case if were looking for a kinematical identification between the initial and final Hilbert spaces as in sec.~\ref{subsec:hs}. In fact, as shown in the cosmological example in ref.~\cite{Agullo:2015qqa}, $\left[ J(t_f) -  J(t_i)\right]$ turns out not to be Hilbert-Schmidt. The latter in particular implies that in the infinite dimensional setting, for arbitrary foliation of the background spacetime, the connection $\cA$ in 
  eq.~\ref{eq:connection} is not unitary. It is only that the dynamical evolution by our Hamiltonian, only a part of which is $\cA$, is unitary. This agrees with the insight of ref.~\cite{Agullo:2015qqa} where it was phrased by saying that only the {\it dynamical} automorphisms of canonical operator algebra are unitarily implementable. 
  
  Finally we would like to make some remarks regarding the shortcomings of our study. Clearly, it would be important to consider the infinite dimensional language from the start, as it was in ref.~\cite{Agullo:2015qqa}. Although phrasing the problem in the language of finite dimensions and taking the limit to infinity is useful and physically insightful, there are issues related to operator domains in Cauchy completions etc. which arise in infinite dimensions that our analysis is completely oblivious of. We leave this problem for those who are better equipped than us in the functional analytical issues and mathematical rigor (see, for e.g., ref.~\cite{1981RSPSA.378..119W}). Another important restriction is that, following
   refs.~\cite{Witten:1993ed, Axelrod:1989xt}, our study was limited to translation invariant complex structures. While this covers some known cases
    (see, for e.g., ref.~\cite{Ashtekar:1975zn}), it certainly is not the most general situation. Last but not least is the fact that the connection $\cA$ is projectively flat. Its curvature
    is $F = -\dfrac{1}{8}\d J^{\ubar{i}}_{~\bar{k}} \wedge \d J^{\bar{k}}_{~\ubar{i}}$. Finiteness of curvature looks like the infinitesimal form of Hilbert-Schmidt condition. The connection can be made flat by adding the so called metaplectic correction. In this paper, we have
    not considered this. It would be interesting to understand the metaplectic correction and its dynamical consequences in the context of generalized unitarity.

\begin{acknowledgments}
The work of MMK was supported  by the grant provided by  Region Bourgogne-Franche-Comte under the program "Stages Monde". 

AM acknowledges useful discussions with Rafi Alam and Sk. Noor Alam.

\end{acknowledgments}

\appendix

\section{Normalization of the state}
\label{app:norm}
State to be normalized:
\begin{align}
 \psi =  \mathscr{N} \, e^{-\frac{i}{2} \o_{\bar{i}\ubar{j}}\varphi^{\bar{i}} \varphi^{\ubar{j}}} \, e^ {\frac{i}{2} \o_{\bar{i}\ubar{j}}\chi^{\bar{i}}_{~\ubar{k}}\varphi^{\ubar{k}} \varphi^{\ubar{j}}}.
\end{align}
 From eq.~\ref{eq:inprod} we have the normalization condition,
 \begin{align}
 1 = |\mathscr{N}|^2  \int \left(\prod_{i=1}^{2N} \, d\varphi^{i} \right) \sqrt{\det{\left(\frac{\o}{2\pi}\right)}}\, \psi^{\ast} \psi.
 \end{align}
 By explicitly inserting the holomorphic projections of eq.~\ref{eq:proj} in $\psi$ we get
 \begin{align*}
 \psi = \mathscr{N} \, e^{-\frac14 \varphi^i \left(\o J_2\right)_{ij} \varphi^j} \, e^{-\frac{i}{4} \varphi^i \left( \o \chi\right)_{ij} \varphi^j - \frac{1}{4} \varphi^i \left( \o \chi J_2\right)_{ij} \varphi^j }.
 \end{align*}
 Hence for $|\psi|^2$ we get
 \begin{align*}
 |\psi|^2 = |\mathscr{N}|^2   e^{-\frac12 \varphi^i \left(\o J_2\right)_{ij} \varphi^j} \, e^{ - \frac{1}{2} \varphi^i \left( \o \chi J_2\right)_{ij} \varphi^j }.
 \end{align*}
 The normalization equation has then just a Gaussian integral,
 \begin{align*}
  1 = |\mathscr{N}|^2  \int \left(\prod_{i=1}^{2N} \, d\varphi^{i} \right) \sqrt{\det{\left(\frac{\o}{2\pi}\right)}}\, e^{-\frac12 \varphi^i M_{ij} \varphi^j},
 \end{align*}
 where $M$ is the matrix $M = \o \left(1 + \chi \right) J_2$. The $\det$ factor in the above integral can come out of the integral since our $\o$ is translation invariant. The remaining
 Gaussian integral gives $\frac{1}{\sqrt{\det \left(M/2\pi\right)}}$. A part of this cancels the $\sqrt{\det \left(\o/2\pi\right)}$ sitting outside and we are left with,
 \begin{align*}
 1 =  |\mathscr{N}|^2 \frac{1}{\sqrt{\det \left[(1 + \chi)J_2\right]}}
 \end{align*} 
Now, the matrix $\left[(1 + \chi)J_2\right]$  can be decomposed in the holomorphic and anti-holomorphic parts with respect to $J_2$, and noting that $\chi$ has only mixed indices, we have
\begin{align}
 \det \left[(1 + \chi)J_2\right] &= \det \left[  \begin{pmatrix} -i\delta^{\bar{i}}_{~\bar{j}} & 0 \\ 0 & i\delta^{\ubar{i}}_{~\ubar{j}} \end{pmatrix}
 +  \begin{pmatrix} 0 & i \chi^{\bar{i}}_{~\ubar{j}} \\ - i\chi^{\ubar{i}}_{~\bar{j}}  & 0 \end{pmatrix} \right] \\
  &= \det\left[\mathbbm{1} - \chi^2 \right],
\end{align}
where the matrix $\chi^2 := \chi^{\bar{i}}_{~\ubar{j}} \chi^{\ubar{j}}_{~\bar{k}}$. Note that $\chi^{\bar{i}}_{~\ubar{j}}$ and $\chi^{\ubar{i}}_{~\bar{j}}$ are complex conjugate matrices therefore $\chi^2$ is  positive.

Now using that $\ln \det = \tr \ln$, we get
 \begin{align}
  \det\left[\mathbbm{1} - \chi^2 \right] &= e^{\tr \ln \left[\mathbbm{1} - \chi^2 \right]} \\
    &< e^{- \tr\left(\chi^2\right) }.
 \end{align}
 Hence the normalization $\mathscr{N}$ is finite iff $\tr \left(\chi^2\right)$ is finite. This is the case iff 
 \begin{align}
 \left(J_2 - J_1\right)^{\bar{i}}_{~\ubar{k}}  \left(J_2 - J_1\right)^{\ubar{k}}_{~\bar{i}} < \infty,
 \end{align}
which is the Hilbert-Schmidt condition.

\bibliography{FoliationsAndUnitarity}
\end{document}